\documentclass[runningheads]{llncs}
\usepackage{graphicx}

\graphicspath{ {figures/} }

\usepackage{wrapfig}

\usepackage{booktabs}

\usepackage{amsfonts}
\usepackage{amsmath}

\usepackage{subfigure}
\usepackage{capt-of}

\usepackage[T1]{fontenc}
\usepackage[utf8]{inputenc}

\usepackage{paralist}

\usepackage{xcolor}

\usepackage[
    breaklinks,
    colorlinks=true,
    urlcolor=blue,
    linkcolor=violet,
    citecolor=violet,
]{hyperref}
\usepackage[capitalise,nameinlink]{cleveref}

\usepackage{xurl}

\usepackage{algorithm}  
\usepackage[noend]{algpseudocode}  

\usepackage{microtype}

\usepackage{orcidlink}
\renewcommand{\orcidID}{\orcidlink}

\usepackage[side,flushmargin]{footmisc}

\let\svthefootnote\thefootnote
\newcommand\freefootnote[1]{%
  \let\thefootnote\relax%
  \footnotetext{#1}%
  \let\thefootnote\svthefootnote%
}

\begin{document}
%
\title{Plan-based Job Scheduling for Supercomputers with Shared Burst Buffers}
\titlerunning{Burst Buffer Aware Plan-based Scheduling}
%
\author{Jan Kopanski \orcidID{0000-0002-3902-6879} \and
Krzysztof Rzadca \orcidID{0000-0002-4176-853X}}
\authorrunning{J. Kopanski and K. Rzadca}
%
\institute{Institute of Informatics, University of Warsaw\\
Stefana Banacha 2, 02-097 Warsaw, Poland\\
\email{jan@kopanski.eu} \quad \email{krzadca@mimuw.edu.pl}}
\maketitle              
\begin{abstract}
\freefootnote{Preprint of the paper accepted at the 27th International European Conference on Parallel and Distributed Computing (Euro-Par 2021), Lisbon, Portugal, 2021, DOI: 10.1007/978-3-030-85665-6\_8}
The ever-increasing gap between compute and I/O performance in HPC platforms, together with the development of novel NVMe storage devices (NVRAM), led to the emergence of the burst buffer concept---an intermediate persistent storage layer logically positioned between random-access main memory and a parallel file system. Despite the development of real-world architectures as well as research concepts, resource and job management systems, such as Slurm, provide only marginal support for scheduling jobs with burst buffer requirements, in particular ignoring burst buffers when backfilling. We investigate the impact of burst buffer reservations on the overall efficiency of online job scheduling for common algorithms: First-Come-First-Served (FCFS) and Shortest-Job-First (SJF) EASY-backfilling. We evaluate the algorithms in a detailed simulation with I/O side effects. Our results indicate that the lack of burst buffer reservations in backfilling may significantly deteriorate scheduling. We also show that these algorithms can be easily extended to support burst buffers. Finally, we propose a burst-buffer--aware plan-based scheduling algorithm with simulated annealing optimisation, which improves the mean waiting time by over 20\% and mean bounded slowdown by 27\% compared to the burst-buffer--aware SJF-EASY-backfilling.

\end{abstract}
%
%
%
\section{Introduction}
With the deployment of Fugaku \cite{dongarra2020report}, supercomputing has already exceeded the threshold of exascale computing. However, this significant milestone only emphasised the challenge of a constantly growing performance gap between compute and I/O~\cite{io-bottlenec}. HPC applications typically alternate between compute-intensive and I/O-intensive execution phases~\cite{7161586}, where the latter is often characterised by emitting bursty I/O requests. Those I/O spikes produced by multiple parallel jobs can saturate network bandwidth and eventually lead to I/O congestion, which effectively stretches the I/O phases of jobs resulting in higher turnaround times. The development of novel storage technologies, such as NVRAM, paves the way to solve the issue of bursty I/O by introducing burst buffers~\cite{6232369}.


A burst buffer is an intermediate, fast and persistent storage layer, which is logically positioned between the random-access main memory in compute nodes and a parallel file system (PFS) in a far (backend) storage. Burst buffers absorb bursty I/O requests and gradually flush them to the PFS, which facilitates, e.g., more efficient checkpointing, data staging or in-situ analysis~\cite{osti_1328312}.

Many recent supercomputers were equipped with burst buffers, including the leading \href{https://www.top500.org/lists/top500/2020/11/}{TOP500} machines:
Fugaku \cite{fugaku}, Summit, and Sierra \cite{10.1109/SC.2018.00055}. In Summit and Sierra, storage devices are installed locally in each compute node -- the \emph{node-local} architecture. The main advantages of this architecture are (a) the linear scaling of I/O bandwidth; and (b) exclusive, consistent and predictable access of jobs to the storage devices, which results in lower variation in I/O performance. An alternative is the \emph{remote shared} architecture (in, e.g., Cori \cite{osti_1393591} and Trinity \cite{osti_1367056}). There are several sub-types of remote shared architectures depending on the placement of burst buffers, which may be located in (1) selected compute nodes, (2) specialised I/O nodes, (3) specialised network-attached storage nodes or (4) in a backend storage system \cite{osti_1328312}. Compared to node-local architectures, shared architectures provide data resiliency and longer residency times, making them more suitable for data staging. An additional benefit of specialised network-attached storage nodes (3) is a transparent maintenance of the supercomputing cluster---replacement of a storage device can be performed without causing downtime of any compute or I/O node. Our paper focuses on shared burst buffer architectures (and our results can be applied to any of the sub-type listed above) because, as we will show later, shared burst buffers challenge the efficiency of modern HPC schedulers.

HPC platforms are managed by middleware, Resources and Jobs Management Systems (RJMS) such as \href{https://slurm.schedmd.com/overview.html}{Slurm}~\cite{10.1007/10968987_3}, responsible for scheduling and executing jobs submitted by users. 
A user submitting a job specifies the binary with all the arguments, but also parameters used for scheduling: the number of requested nodes and the walltime -- an upper bound on the processing time.
RJMSs usually implement some sort of backfilling algorithm for scheduling \cite{1374857}. Perhaps the most common is the First-Come-First-Served (FCFS) EASY-backfilling (aggressive backfilling) \cite{10.1007/3-540-36180-4_4}, in which jobs are queued according to their arrival time (sometimes weighted by the job's or the user's priority, but this is orthogonal to our approach). Another common scheduling policy is the Shortest-Job-First (SJF; SPF) EASY-backfilling \cite{carastansantos:hal-02237895} that sorts pending jobs by their walltime. SJF usually improves user-centric metrics such as mean waiting time or mean slowdown. Regardless of the ordering, the EASY-backfilling algorithm may backfill (execute a job out of order) whenever there are enough free nodes, and the backfilled job would not delay the job at the head of the queue.


The current approach to managing burst buffers by RJMS seems to just extend the user-provided job description by the requested volume of burst buffer storage. This requested volume is then granted for the job's entire duration (just as compute nodes are granted for the entire duration). However, when we analysed backfilling in Slurm, we noticed that these burst buffer requests are treated quite differently from compute nodes. \href{https://slurm.schedmd.com/burst_buffer.html}{Slurm documentation} explains this phase as: ``After expected start times for pending jobs are established, allocate burst buffers to those jobs expected to start earliest and start stage-in of required files.'' It means that the burst buffers are allocated after backfilling, which may starve jobs requiring burst buffers. 


\paragraph{Contributions.}
We address the problem of efficient online job scheduling in supercomputers with shared burst buffer architecture as follows:
\begin{enumerate}
\item By simulation,
  we show that in the EASY-backfilling algorithm, the Slurm-like decoupling of burst buffer reservations from processor reservations leads to heavy-tailed distribution turnaround times.
    \item We point to a possible starvation issue of jobs with burst buffer requirements in Slurm backfilling implementation.
    \item We show a simple extension of EASY-backfilling which significantly improves scheduling efficiency by considering both burst buffers and processors.
    \item We propose a burst-buffer--aware plan-based scheduling algorithm with simulated annealing optimisation, which improves the mean waiting time by over 20\% and mean bounded slowdown by 27\% compared to the SJF EASY-backfilling with burst buffer reservations.
\end{enumerate}



\noindent Source code associated with this paper is available at:\\
\url{https://github.com/jankopanski/Burst-Buffer-Scheduling}









\section{Related work} \label{sec:related-work}
Scheduling in HPC is a vast research area. Below, we review only the papers addressing the problem of scheduling with burst buffer resources.

FCFS and EASY-backfilling are examples of queue-based scheduling algorithms. Queue-based schedulers periodically attempt to launch jobs based on only the current state of a system. An alternative is plan-based scheduling which creates an execution plan of all pending jobs~\cite{hovestadt2003scheduling}. Zheng \textit{et al.} \cite{7776518} used simulated annealing for finding an optimal execution plan. However, they only considered standard, CPU jobs. 
Our proposed plan-based algorithm (\cref{sec:algorithm-plan}) extends their approach with burst buffer requirements and introduces several improvements to their simulated annealing optimisation, significantly reducing the number of iterations required to find a near-optimal solution. 

Efficient job scheduling in platforms with shared burst buffer architecture was studied by Lackner, Fard and Wolf \cite{8752797}.
A job waiting for burst buffers may end up with a higher turnaround time than if it started earlier, but without fast persistent storage (which could increase its runtime).
\cite{8752797} proposed an extension to backfilling by estimating job's turnaround times with and without the access to burst buffers.
In general, their solution is complementary to ours and could be incorporated into our plan-based scheduling.

Fan \textit{et al.} \cite{10.1145/3307681.3325401} formulated the multi-resource scheduling problem as a multi-objective integer linear program (ILP), which maximises both compute and storage utilisation. The ILP is optimised by a genetic algorithm generating a Pareto front. To limit the computational complexity, they perform the optimisation only on a fixed-size window at the front of a waiting queue.
A potential problem in their algorithm may arise in the step of resource allocation. 
For instance, in a Dragonfly cluster topology, we prefer to allocate nodes for a job within a single group. Such topology-awareness is not readily reflected in the proposed ILP. In contrast, our plan-based approach is as generic as backfilling and thus does not impose any limitations on resource allocation.

Herbein \textit{et al.} \cite{10.1145/2907294.2907316} considered I/O contention between burst buffers and PFS. They modified EASY-backfilling by allowing it to start only those jobs that do not oversaturate network links and switches. While this change solves the issue of stretching I/O phases, it might result in underutilisation of the network bandwidth. Consequently, although the job running time is minimised, the waiting time may increase, leading to a higher turnaround time. 


A hybrid solution that combines detailed information from RJMS with low-level I/O requests scheduling was proposed by Zhou \textit{et al.} \cite{7307592}. They presented an I/O-aware scheduling framework for concurrent I/O requests from parallel applications. 
It enables the system to keep the network bandwidth saturated and keep a uniform stretch of I/O phases of jobs, improving average job performance. In that sense, it is an alternative solution to Herbein \textit{et al.} However, as Zhou \textit{et al.} approach does not require any modification of the main scheduling algorithm, it can be incorporated into our burst-buffer--aware plan-based scheduling.

\section{Scheduling algorithms} \label{sec:algorithms}
\subsection{Example} \label{sec:example}

\begin{figure}[p]
  \centering
  \subfigure{\includegraphics[page=1,width=0.49\linewidth]{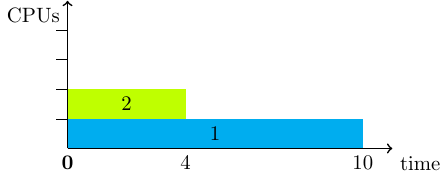}}
  \subfigure{\includegraphics[page=2,width=0.49\linewidth]{starvation.pdf}}
  \subfigure{\includegraphics[page=3,width=0.49\linewidth]{starvation.pdf}}
  \subfigure{\includegraphics[page=4,width=0.49\linewidth]{starvation.pdf}}
  \subfigure{\includegraphics[page=5,width=0.49\linewidth]{starvation.pdf}}
  \subfigure{\includegraphics[page=6,width=0.49\linewidth]{starvation.pdf}}
  \caption{FCFS EASY-backfilling \emph{without} future burst buffer reservations may delay all jobs in a queue behind the first job (3), which received a reservation of processors. Job 3 cannot start at its scheduled time (t=4min, bottom-left) as there are not enough burst buffers until job 1 completes. Note that subfigures show the schedule in subsequent time moments (t=0 min top-left, t=1 top-right, t=2 middle-left, etc.)}
  \vspace{0.5cm}
  \label{fig:example-starvation}
  \subfigure{\includegraphics[page=1,width=0.49\linewidth]{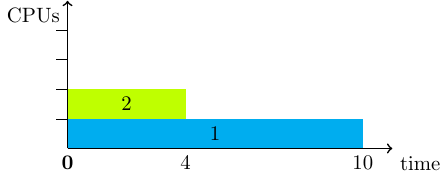}}
  \subfigure{\includegraphics[page=2,width=0.49\linewidth]{bb-aware.pdf}}
  \subfigure{\includegraphics[page=3,width=0.49\linewidth]{bb-aware.pdf}}
  \subfigure{\includegraphics[page=4,width=0.49\linewidth]{bb-aware.pdf}}
  \subfigure{\includegraphics[page=5,width=0.49\linewidth]{bb-aware.pdf}}
  \subfigure{\includegraphics[page=7,width=0.49\linewidth]{bb-aware.pdf}}
  \caption{Schedule by FCFS EASY-backfilling \emph{with} burst buffer reservations. Job 3 is scheduled after job 1 completes, which permits backfilling all the remaining jobs.}
  \label{fig:example-bb-aware}
\end{figure}

\begin{wraptable}{r}{6.5cm}
  \vspace{-1cm}
  \centering
  \caption{Example of jobs with an inefficient schedule with the standard EASY-backfilling.}
  \begin{tabular}{ccccc}
    \toprule
    Job & Submit[m] & Runtime[m] & CPU & BB[TB]\\
    \midrule
    1 & 0 & 10 & 1 & 4\\
    2 & 0 & 4 & 1 & 2\\
    3 & 1 & 1 & 3 & 8\\
    4 & 2 & 3 & 2 & 4\\
    5 & 3 & 1 & 3 & 4\\
    6 & 3 & 1 & 2 & 2\\
    7 & 4 & 5 & 1 & 2\\
    8 & 4 & 3 & 2 & 4\\
    \bottomrule
  \end{tabular}
  \label{tab:jobs}
  \vspace{-0.7cm}
\end{wraptable}

We start with an illustrative example of how a burst buffer reservation may impact the FCFS EASY-backfilling schedule. Suppose we are executing jobs as defined by \cref{tab:jobs} on a small cluster with 4 nodes (denoted by CPUs) and 10 TB of shared burst buffer storage. 
For simplicity, walltime of each job is equal to its runtime (perfect user runtime estimates). The BB column denotes the total amount of burst buffer requested by a job. The scheduler runs periodically at every minute. The first iteration of scheduling starts at time 0 and also includes jobs submitted at time 0. 
For comparison of algorithms, we present how this set of jobs would be scheduled by EASY-backfilling without (\cref{fig:example-starvation}) and with (\cref{fig:example-bb-aware}) burst buffer reservations.
Filled rectangles represent started jobs; stripped represent future FCFS reservations.


Let us focus our attention on job 3. 
\cref{fig:example-starvation} shows the schedule created by the backfilling without burst buffer reservations. At the time 0 (top-left), two jobs are submitted and started immediately. After a minute (top-right), job 3 is submitted, but there are not enough available processors to start it. Concerning only CPUs, jobs 1 and 3 would be able to run in parallel after the completion of job 2. However, their summarised burst buffer requirements exceed the total capacity of 10 TB available in the cluster. As the standard EASY-backfilling does not consider burst buffers in scheduling, it would schedule job 3 just after job 2. At time 2 (middle-left), job 4 is submitted. Despite having enough free processors and burst buffers, it cannot be launched instantly as it would delay job 3, which is at the front of the waiting queue and hence received a reservation of 3 CPUs for the time period 4-5. The same situation applies to job 5, which given the resources, could be backfilled in place of job 3, but would then delay its scheduled execution. It is not the case for job 6, which can be backfilled before job 3. An unusual situation happens at time 4 (bottom-left). Job 3 does not have enough burst buffers to start the execution and prevents all other jobs from starting. It causes most of the processors in the cluster to remain idle until the completion of job 1. Indeed, job 3 behaves like a barrier in the scheduling of other jobs. The last time step presents the moment just after the completion of job 1. Job 3 is finally able to start. Additionally, job 7 is backfilled ahead of jobs 4 and 5 as it has enough processors and burst buffers to start immediately.

\begin{figure}[tb]
  \setlength\abovecaptionskip{-1.4\baselineskip}
  \centering
  \includegraphics[width=\linewidth]{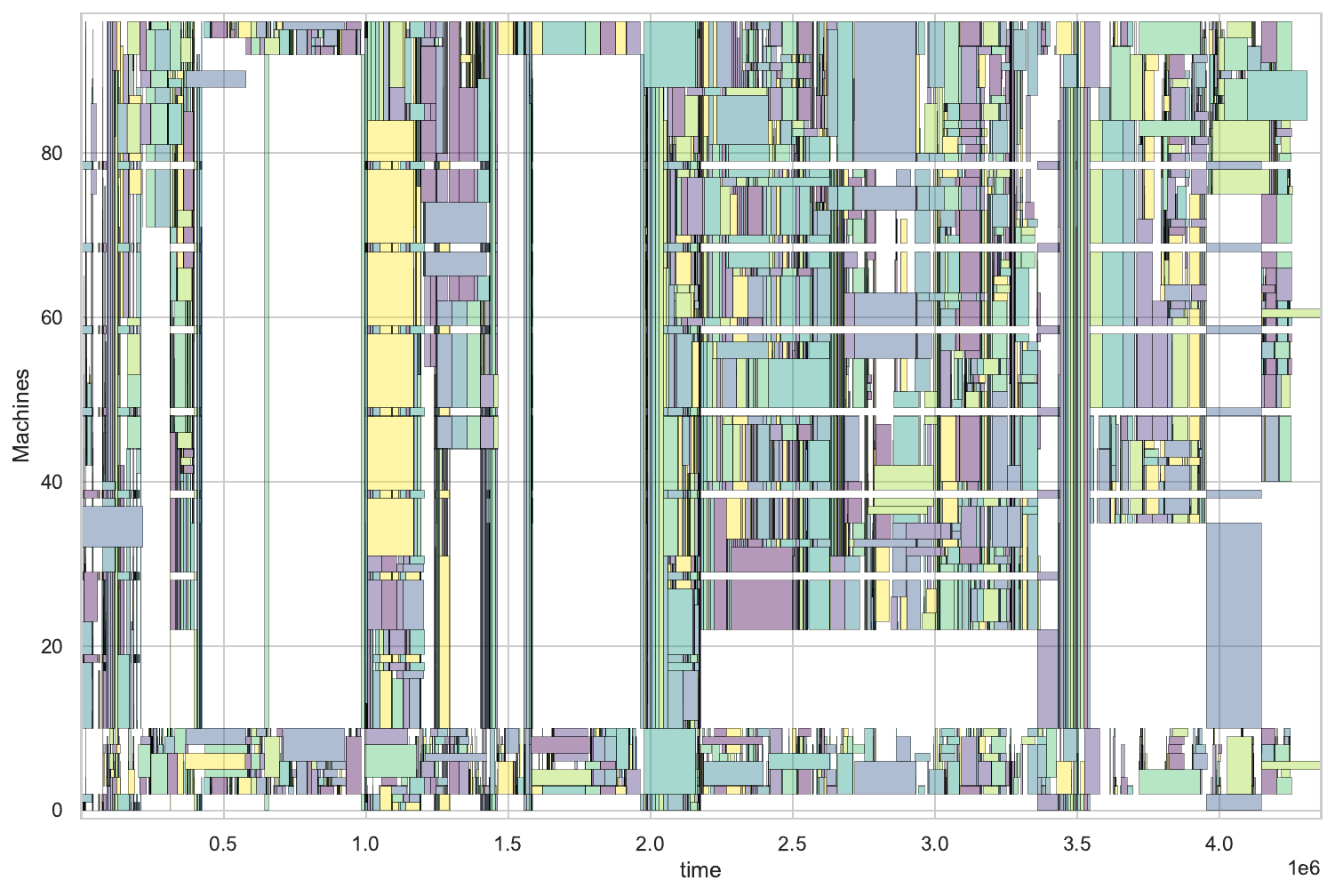}
  \caption{Gantt chart of the first 3500 jobs executed by FCFS EASY-backfilling without future reservations of burst buffers. The large empty spaces in the chart, which are immediately followed by tall jobs, indicate significantly underutilised compute resources.}
  \label{fig:gantt}
\end{figure}

\cref{fig:example-bb-aware} shows how FCFS EASY-backfilling \emph{with burst buffer reservations} schedules the same set of jobs. The first difference from the previous algorithm appears at time step 1. Although the newly arrived job 3 cannot start at that time, its burst buffer requirements are recognised and it is scheduled after job 1 completes. Thus, job 4 can start immediately when submitted, as it will not delay job 3. At time 4, all the new jobs 5, 6, 7, 8 can be backfilled ahead of job 3. Lastly, at time 10, there are only left jobs 3 and 7.

Although this example may seem artificial, it distils the problem we observed when simulating a larger set of jobs (see \cref{fig:gantt}).

\subsection{Burst buffer reservations in backfilling}\label{sec:bb-backfill}
In this section, we show how to extend EASY-backfilling with burst buffer reservations. 
\cref{alg:backfill} shows the pseudocode of EASY-Backfilling. Jobs in the waiting queue $Q$ are stored in the order of arrival. At the beginning, the algorithm attempts to launch jobs according to FCFS policy until the first job $J$ for which there are not enough free resources (processors or burst buffers). The reservation for processors is acquired for $J$ at the earliest possible time in the future. Next, the remaining jobs in $Q$ are backfilled ahead of $J$ under the condition that they would not delay the scheduled execution of $J$.

The standard EASY-Backfilling does not include the part presented in the square brackets (line \ref{alg:backfill:reservation}). This minor change---a future reservation of burst buffers together with processors prior to the backfilling process---solves the problem presented in the example in \cref{sec:example}. 


\newlength{\oldtextfloatsep}\setlength{\oldtextfloatsep}{\textfloatsep}
\begin{algorithm}[!b]
\caption{EASY-Backfilling scheduling}
\label{alg:backfill}
\begin{algorithmic}[1]
\Procedure{FCFS}{$Q$}
\Comment{$Q$ - queue of pending jobs}
  \For{$J \in Q$}
    \If{there are enough free processors and burst buffers for $J$}
        \State Launch $J$ and remove it from $Q$
        \Comment{Allocations cannot overlap}
    \Else
        \State Break
    \EndIf
  \EndFor
\EndProcedure
\Procedure{Backfill}{$Q$}
\label{alg:backfill:fill}
  \For{$J \in Q$}
  \Comment{Allocations cannot overlap with reservations}
    \If{there are enough free processors and burst buffers for $J$}
        \State Launch $J$ and remove it from $Q$
    \EndIf
  \EndFor
\EndProcedure
\Procedure{EASY-Backfilling}{$Q$}
  \State \Call{FCFS}{$Q$}
  \State $J \gets$ pop the first job from $Q$
  \State Reserve compute [and storage] resources for $J$ at the earliest time in the future \label{alg:backfill:reservation}
  \If{SJF}
  \State Sort $Q$ ascending by walltime
  \EndIf
  \State \Call{Backfill}{$Q$}
  \State Remove reservations for $J$
  \Comment{Will be reacquired in the next scheduling}
  \State Push back $J$ at the front of $Q$
\EndProcedure
\end{algorithmic}
\end{algorithm}

In principle, Slurm implements conservative backfilling. However, there is an issue in Slurm implementation of burst buffer support, which could possibly lead to starvation of jobs with burst buffer requirements. 
Slurm allows to delay a job requesting burst buffer if it has not started a stage-in phase. In this case, the job does not receive a reservation of processors. Therefore, other jobs can be backfilled ahead of it. In our experimental workload, every job requires burst buffers, thus every job can be arbitrarily delayed. 
Assuming that all jobs require burst buffers and each job is executed right after the stage-in phase, as in our model (\cref{fig:job-model}),
Slurm backfilling works similarly to backfilling without any future reservations---procedure Backfill at line \ref{alg:backfill:fill} (but without the reservation in line~\ref{alg:backfill:reservation}).
There are still some differences between Slurm scheduling and the Backfill procedure. First, Slurm decouples allocation of burst buffers and allocation of processors. Slurm also has an FCFS scheduling loop running concurrently to backfilling.

In conclusion, the large empty spaces visible in \cref{fig:gantt} only happen in the theoretical EASY-backfilling without future reservations of burst buffers. The current, greedy scheduling in Slurm solves this but at the expense of potentially arbitrary delays of jobs with burst buffer requirements. A simple extension of backfilling with future simultaneous processor and burst buffer reservations eliminates both problems.

\subsection{Plan-based scheduling} \label{sec:algorithm-plan}
The general idea of plan-based scheduling is to create an execution plan for all jobs in the waiting queue. An execution plan is an ordered list of jobs with their scheduled start times and assigned resources. Naturally, only a few jobs could be launched immediately. Others are provided with non-overlapping reservations of resources at different points of time in future for the duration of their walltimes. The easiest way to create the execution plan is to iterate over the queue of jobs and for each job find the earliest point in time when sufficient resources are available. The quality of the obtained plan depends on the order of jobs. Therefore, it is possible to define an optimisation objective and perform a search over permutations of jobs.

\begin{algorithm}[tb]
\caption{Plan-based scheduling}
\label{alg:plan}
\begin{algorithmic}[1]
\Procedure{Plan-Based}{$Q$, $r$, $N$, $M$}
  \If{$Q$ is small ($\leq 5$ jobs)}
    \State Perform exhaustive search over all permutations
    \State Reserve resources; Launch jobs; Return
  \EndIf
  \State $P_{\text{best}}, S_{\text{best}} \gets$ find a permutation with the lowest score among  candidates
  \State $P_{\text{worst}}, S_{\text{worst}} \gets$ find a permutation with the highest score among  candidates
  \If{$S_{\text{best}} \neq S_{\text{worst}}$}
    \Comment{Simulated annealing}
    \State $T \gets S_{\text{worst}} - S_{\text{best}}$ \label{alg:plan:temp}
    \Comment{Initial temperature}
    \State $P \gets P_{\text{best}}$
    \For{$N$ iterations}
      \For{$M$ iterations}
        \State $P' \gets$ swap two jobs at random positions in $P$\
        \State Create execution plans for $P$ and $P'$
        \Comment{Details in \cref{sec:algorithm-plan}}
        \State $S \gets$ calculate score for $P$ based on the execution plan
        \State $S' \gets$ calculate score for $P'$ based on the execution plan
        \If{$S' < S_{\text{best}}$}
          \State $S_{\text{best}} \gets S'$; $P_{\text{best}} \gets P'$
          \State $S \gets S'$; $P \gets P'$
        \ElsIf{$S' < S \lor \text{random}(0,1) < e^{(S - S')/T}$}
          \State $S \gets S'$; $P \gets P'$
        \EndIf
      \EndFor
      \State $T \gets r \cdot T$
      \Comment{Temperature cooling}
    \EndFor
  \EndIf
  \State Launch jobs and reserve resources according to the execution plan of $P_{\text{best}}$
\EndProcedure
\end{algorithmic}
\end{algorithm}

We further investigate and extend the plan-based approach to scheduling with simulated annealing optimisation proposed by Zheng \textit{et al.} \cite{7776518}. First, we extend the reservation schema with the reservations for burst buffer requests. Second, we apply several modifications to the simulated annealing. Our plan-based scheduling is presented in \cref{alg:plan}. The input parameters $Q$, $r$, $N$, $M$ denote the waiting queue, cooling rate, number of cooling steps and constant temperature steps respectively.

As the optimisation objective, we minimise a sum of waiting times of jobs weighted by an exponent $\alpha \in \mathbb{R}^+$ (definition of the waiting time is given in \cref{fig:job-model}). For a permutation $P$, let $W_j^P$ denote the waiting time of the $j$-th job in $Q$ according to the execution plan created based on $P$.
\begin{equation}
    \min_P \sum_{j \in Q} (W_j^P)^\alpha
    \label{eq:objective}
\end{equation}
We denote the sum in the above formula as a score $S$ of the execution plan of $P$. Zheng \textit{et al.} observed that for small values, such as $\alpha=1$, this objective allows one job to be significantly postponed in favour of other jobs as long as the total waiting time of all jobs improves, which eventually may even lead to starvation. The higher is the value of $\alpha$, the more penalised is the objective function for delaying a job. Plan-based scheduling does not ensure as strict fairness criteria as reservations in EASY-backfilling, but increasing the value of $\alpha$ may be perceived as a soft mechanism for ensuring fair resource sharing. 

We enhance the simulated annealing algorithm of~\cite{7776518} as follows.
First, we introduce a set of candidates $I$ for the initial permutation used to generate the best and worst initial scores ($S_{\text{best}}$, $S_{\text{worst}}$), which allow us to set an optimal initial temperature $T$ according to the method proposed by~\cite{10.1023/B:COAP.0000044187.23143.bd}. We define the set of 9 initial candidates by sorting jobs according to 9 different criteria: (1) the order of submission (FCFS); (2) ascending and (3) descending by the requested number of processors, (4) ascending and (5) descending by the requested size of burst buffer per processor, (6) ascending and (7) descending by the ratio of the requested size of burst buffer per processor to the requested number of processors, (8) ascending and (9) descending by walltime. 

Second, we make the search space exploration faster by (1) introducing exhaustive search over all possible permutations for small queues (up to 5 jobs); (2) faster cooling ($r=0.9$, $N=30$, $M=6$); and (3) skipping the annealing if it is unlikely to find a better permutation---when the scores of the best and the worst initial candidates are the same ($S_{\text{best}} = S_{\text{worst}}$). As a result, our algorithm finds a near-optimal permutation in only $N \cdot M + |I| = 189$ iterations, compared to $\lceil 100 \log_{0.9}(0.0001) \rceil = 8742$ iterations by Zheng \textit{et al.} The number of iterations has principal importance in online job scheduling, where the time available for scheduling is limited.

\section{Simulation} \label{sec:simulation-model}
\subsection{Method}
To evaluate and compare scheduling algorithms, we created a detailed supercomputer simulator. We extended \href{https://batsim.readthedocs.io/en/latest/}{Batsim} \cite{dutot:hal-01333471,poquet:tel-01757245}---a simulator framework of RJMS based on \href{https://simgrid.org/}{SimGrid} \cite{casanova:hal-01017319}, which is capable of simulating I/O contention and I/O congestion effects. We used \href{https://gitlab.inria.fr/batsim/pybatsim}{Pybatsim scheduler} to implement our algorithms and manage I/O side effects.

\paragraph{Platform model:}
Our simulated cluster is modelled according to a \href{https://simgrid.org/doc/latest/Platform_Examples.html#dragonfly-cluster}{Dragonfly topology}. It consists of 108 nodes divided into 3 groups, each group contains 4 chassis, each chassis has 3 routers, and there are 3 nodes attached to each router. However, only 96 of the nodes are compute nodes. The other 12 nodes are assigned a storage role. That is, a single node in every chassis is dedicated to being a burst buffer node. This type of shared burst buffer architecture resembles the architecture of Fugaku, where one of every 16 compute nodes contains SSDs for burst buffers \cite{fugaku}. In our model, a single compute node is equivalent to a single processor/CPU. While our simulated cluster has a limited size (only 108 nodes), we consider it sufficient to show the effects of the proposed scheduling policies. We could not scale to more nodes due to a technical issue with Batsim reporting errors on larger SimGrid Dragonfly clusters.

The bandwidth of a compute network is set to model 10 Gbit/s Ethernet. The compute network is connected with a single shared link to one additional node which represents PFS. We set the bandwidth of this link to 5 GB/s, based on data from the \href{https://www.vi4io.org/}{IO500 list}, to make I/O side effects noticeable.
We set the total burst buffer capacity to the expected total burst buffer request when all nodes are busy, which we calculate based on the average of the fitted log-normal distribution of burst buffer request per processor (described below). We divide this capacity equally among the storage nodes.


\paragraph{Job model:}
\begin{figure}[tb]
  \setlength\abovecaptionskip{-1.2\baselineskip}
  \centering
  \includegraphics[width=\textwidth]{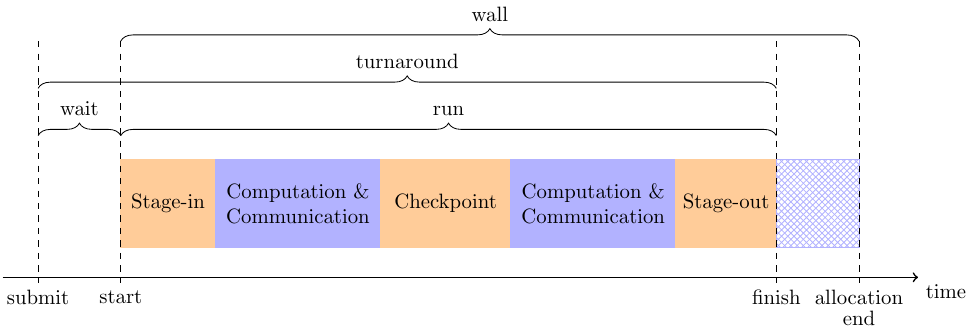}
  \caption{Job execution model used in simulations and job's performance metrics. After the job starts, it get the data from PFS (stage-in), then performs between 1 and 10 computation phases; after each, but last, computation phase, the job checkpoints. After the last computation phase, the job transfers the results to PFS (stage-out).}
  \label{fig:job-model}
\end{figure}
We consider parallel, non-preemptive and rigid jobs. Our simulated job model (\cref{fig:job-model}) includes data staging and checkpointing as burst buffer use cases, which are interleaved with computation and communication phases. A job is divided into a variable number of phases (from 1 to 10) based on the specified number of computing operations. They are interleaved by I/O phases which simulate checkpointing---one of the major use cases of burst buffers. When a job checkpoints, it transfers data from compute nodes to the assigned burst buffer nodes; meanwhile, the job's computations are suspended. After the checkpoint completes, data transfer from burst buffers to PFS is triggered, and the next computation phase starts concurrently. Furthermore, jobs start and complete with data staging phases between PFS and burst buffers.
The size of the data transfers is equal to the requested burst buffer size. 

\paragraph{Workload model:}
In order to perform experiments on a realistic workload, we decided to transform the \href{https://www.cs.huji.ac.il/labs/parallel/workload/l_kth_sp2/index.html}{KTH-SP2-1996-2.1-cln} log from the \href{https://www.cs.huji.ac.il/labs/parallel/workload/}{Parallel Workload Archive} (PWA) \cite{FEITELSON20142967} into the \href{https://batsim.readthedocs.io/en/latest/input-workload.html}{Batsim input format}, resulting in the workload with 28453 jobs. We selected this log as it was recorded on a cluster with 100 nodes, the closest to our simulated cluster. From the log, we extract submission times, walltimes and the requested number of processors. The requested memory size would be especially valuable for modelling burst buffer requests, but this data is not available in this log. We use the log in two different settings: first, we perform some experiments on the complete log; second, we split the workload into 16 non-overlapping, three-week-long parts to measure the variability of our results. 

\paragraph{Burst buffer request model:}
As workloads in PWA do not contain information on burst buffer requests, researchers either collect new logs \cite{10.1145/3307681.3325401} or create probabilistic workload models \cite{8752797,10.1145/2907294.2907316}. As the log from \cite{10.1145/3307681.3325401} was unavailable, we supplemented jobs from the KTH-SP2-1996-2.1-cln log with a probabilistic distribution of burst buffer requests. Under the assumption that a burst buffer request size is equal to the main memory request size, we created a model of burst buffer request size per processor based on the \href{https://www.cs.huji.ac.il/labs/parallel/workload/l_metacentrum2/index.html}{METACENTRUM-2013-3} log. Burst buffer request size equal to the requested RAM size is a representative case for modelling data staging and checkpointing (saving a distributed state of an entire job).

At first, we investigated a cross-correlation between the requested memory size and the number of processors. According to empirical cumulative distribution functions, it appears only for large jobs with at least 64 processors. However, large jobs contribute to only 11\% of processor time. Therefore we model burst buffer requests per processor independently from the job size. We fitted several long-tail distributions to empirical data, which resulted in the best fit achieved by a log-normal distribution. We validated the quality of fitting with 5-fold cross-validation and Kolmogorov-Smirnov D-statistic test.

\subsection{Results} \label{sec:results}
We evaluate and compare the following scheduling policies:
\begin{description}
\item[fcfs] FCFS without backfilling;
\item[fcfs-easy] FCFS EASY-backfilling without reservations for burst buffers;
\item[filler] Perform the backfill procedure from \cref{alg:backfill} but without reserving resources for queued jobs in future (without line \ref{alg:backfill:reservation});
\item[fcfs-bb] FCFS EASY-backfilling with simultaneous reservations for processors and burst buffers;
\item[sjf-bb] SJF EASY-backfilling with simultaneous reservations for processors and burst buffers;
\item[plan-$\alpha$] burst-buffer--aware plan-based scheduling; the number in the policy name (e.g., plan-2) is a value of the parameter $\alpha$ from \cref{eq:objective}.
\end{description}

\begin{figure}[p]
  \centering
  \begin{minipage}[b]{0.49\linewidth}
    \setlength\abovecaptionskip{-1.1\baselineskip}
    \includegraphics[width=\linewidth]{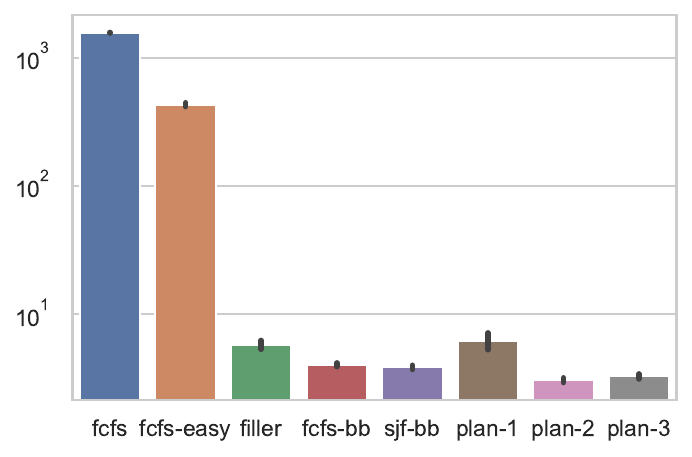}
    \caption{Mean waiting time [hours].}
    \label{fig:waiting-time_mean}
  \end{minipage}
  \begin{minipage}[b]{0.49\linewidth}
    \setlength\abovecaptionskip{-1.1\baselineskip}
    \includegraphics[width=\linewidth]{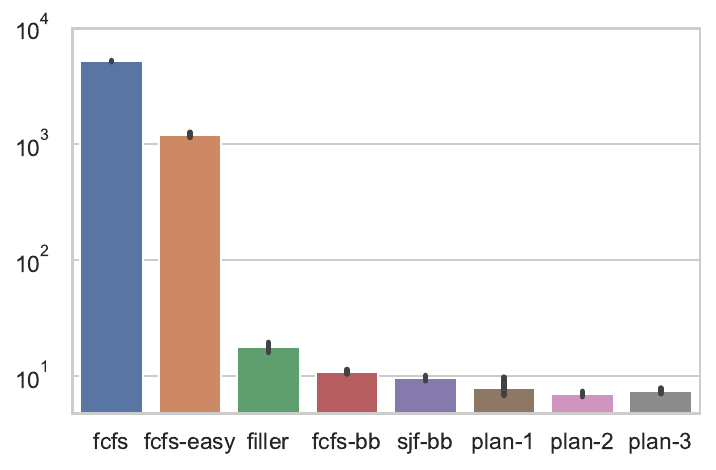}
    \caption{Mean bounded slowdown.}
    \label{fig:bounded-slowdown_mean}
  \end{minipage}
\end{figure}
\begin{figure}[p]
  \centering
  \begin{minipage}[b]{0.49\linewidth}
    \setlength\abovecaptionskip{-1.1\baselineskip}
    \includegraphics[width=\linewidth]{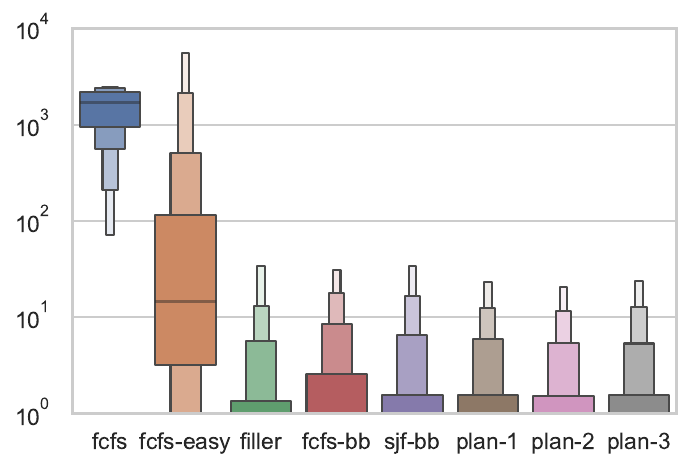}
    \caption{Waiting time quantiles [h].}
    \label{fig:waiting-time_boxen}
  \end{minipage}
  \begin{minipage}[b]{0.49\linewidth}
    \setlength\abovecaptionskip{-1.1\baselineskip}
    \includegraphics[width=\linewidth]{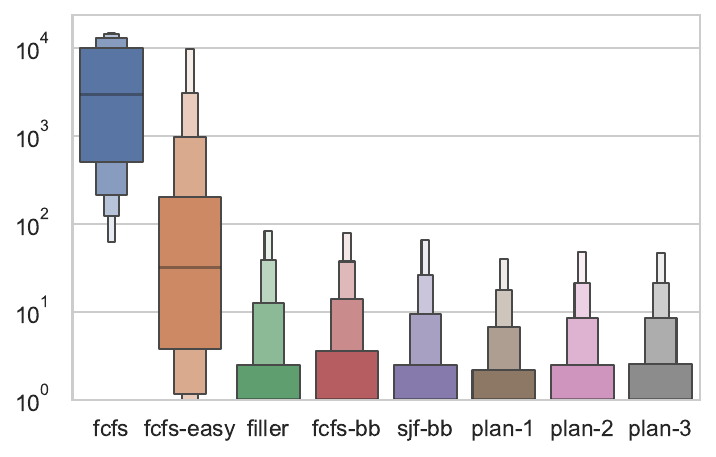}
    \caption{Bounded slowdown quantiles.}
    \label{fig:bounded-slowdown_boxen}
  \end{minipage}
\end{figure}
\begin{figure}[p]
  \centering
  \begin{minipage}[b]{0.49\linewidth}
    \setlength\abovecaptionskip{-1.1\baselineskip}
    \includegraphics[width=\linewidth]{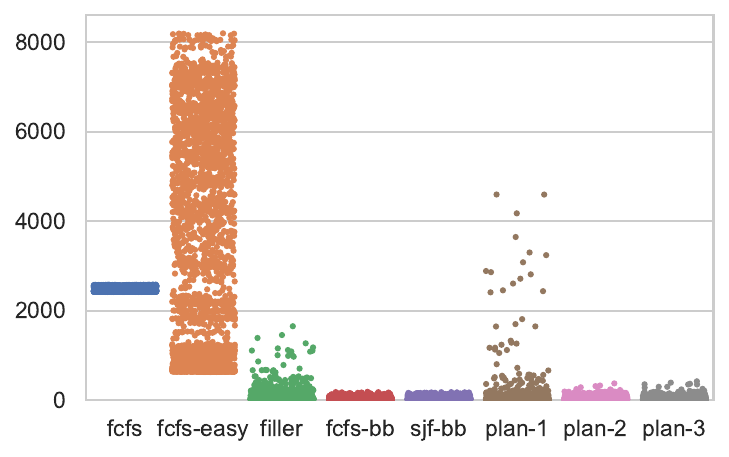}
    \caption{Waiting time tail distribution [h].}
    \label{fig:waiting-time_dist}
  \end{minipage}
  \begin{minipage}[b]{0.49\linewidth}
    \setlength\abovecaptionskip{-1.1\baselineskip}
    \includegraphics[width=\linewidth]{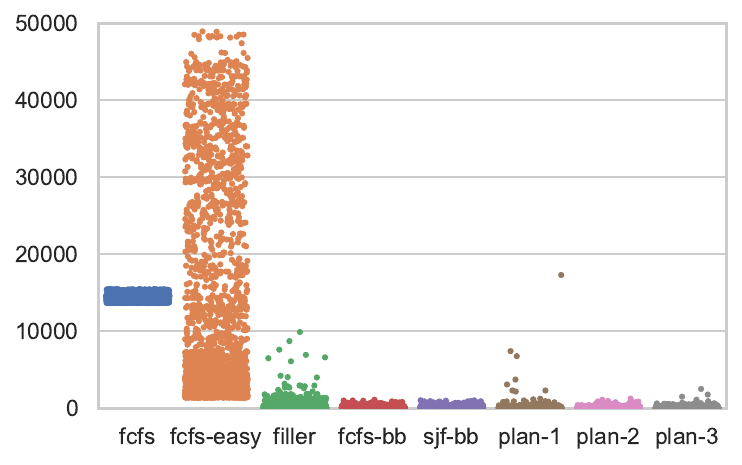}
    \caption{Bounded slowdown tail dist.}
    \label{fig:bounded-slowdown_dist}
  \end{minipage}
\end{figure}
\begin{figure}[p]
  \centering
  \begin{minipage}[b]{0.49\linewidth}
    \setlength\abovecaptionskip{-1.1\baselineskip}
    \centering
    \includegraphics[width=\linewidth]{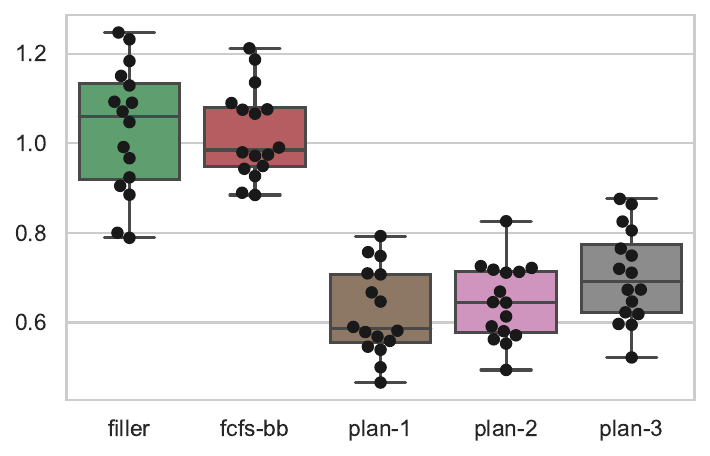}
    \caption{Normalised mean waiting time \mbox{distribution} over 16 workload parts.}
    \label{fig:waiting-time_split}
  \end{minipage}
  \begin{minipage}[b]{0.49\linewidth}
    \setlength\abovecaptionskip{-1.1\baselineskip}
    \centering
    \includegraphics[width=\linewidth]{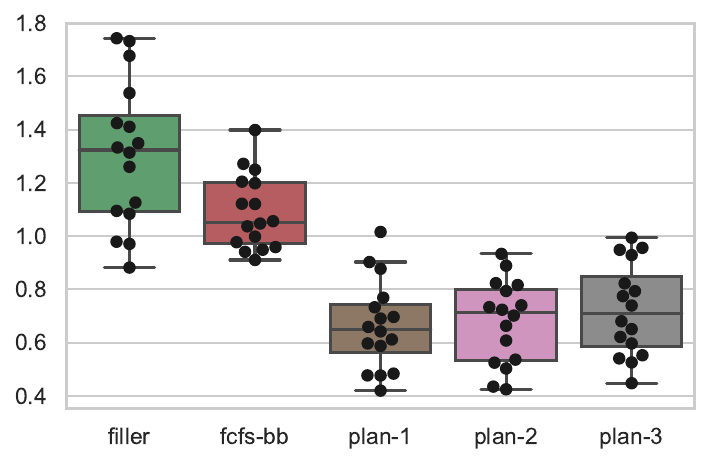}
    \caption{Normalised mean bounded slowdown distribution over 16 workload parts.}
    \label{fig:bounded-slowdown_split}
  \end{minipage}
\end{figure}

\noindent We show mean waiting time (\cref{fig:waiting-time_mean}) and mean bounded slowdown (\cref{fig:bounded-slowdown_mean}, bounded for jobs shorter than 10 minutes). The small black bars in these plots show 95\% confidence intervals.

\cref{fig:waiting-time_mean} and \cref{fig:bounded-slowdown_mean} show the first surprising result: excluding the baseline fcfs, fcfs-easy has \emph{two orders of magnitude higher} mean waiting time and bounded slowdown than other policies.
The simple extension of the FCFS-backfilling to explicitly reserve burst buffers (fcfs-bb) results in significant improvements. When the scheduler is additionally able to change the ordering of the jobs (sjf-bb), the gains are even higher. 
fcfs-easy shows better distribution than fcfs for the majority of jobs. However, it changes for the last 32-quantile. The explanation for this observation is visible in the tail distribution plot \cref{fig:waiting-time_dist}, which for each policy presents individual values for 3000 jobs with the highest waiting times. We see there an extremely significant dispersion of waiting times of fcfs-easy, which also affects the bounded slowdown (\cref{fig:bounded-slowdown_dist}). Overall, fcfs-easy on average achieves better results than fcfs, but it can considerably delay numerous jobs. These results quantitatively confirm the example from \cref{sec:example}. 

The simple filler shows good distribution of waiting times and bounded slowdown, also in the quantile plots (\cref{fig:waiting-time_boxen} and \cref{fig:bounded-slowdown_boxen} are boxenplots, also called letter-value plots \cite{doi:10.1080/10618600.2017.1305277}, with a logarithmic scale). However, the tail distribution plots (\cref{fig:waiting-time_dist} and \cref{fig:bounded-slowdown_dist}) show that filler is likely to disproportionately delay some jobs---pointing to their near-starvation, which we hinted at in \cref{sec:bb-backfill}.

Our proposed plan-based scheduling algorithm, plan-2, outperforms all other policies. As opposed to queue-based scheduling, plan-based scheduling may leave a cluster underutilised at a given moment to enable earlier execution of jobs in the future. Although this behaviour can result in a higher median waiting time, it should improve higher-order quantiles as seen in \cref{fig:waiting-time_boxen}. However, as plan-1 has much flexibility in reordering jobs, it is likely to fully utilise the cluster by taking jobs from any position in the queue for instant execution. This flexibility comes at the cost of significant delay of selected jobs, as shown in \cref{fig:waiting-time_dist}.

For a more robust comparison between the algorithms, we split the workload into 16 three-week-long parts. We compute the average for each part and each policy, and then normalise it by the corresponding average from the sjf-bb policy. We chose sjf-bb as the reference policy, because on the whole workload, its mean waiting time is smaller by 4.5\% than fcfs-bb. Boxplots in \cref{fig:waiting-time_split} and \cref{fig:bounded-slowdown_split} show statistics. Moreover, the 16 dots on each box correspond to the 16 averages, one for each three-week-long part. These results might be interpreted in the following way: the lower the boxplot, the better the scheduling efficiency. This analysis confirms our results on the whole trace: in most cases, filler has a higher mean than non-starving, burst-buffer--aware policies; and plan-based scheduling outperform greedy list scheduling. Additionally, in the shorter workloads, plan-1 outperforms plan-2 because of the limited capability of delaying jobs.



\section{Conclusion} \label{sec:conclusion}
The lack of reservations of burst buffers acquired simultaneously with processors in backfilling can cause significant inefficiency in online job scheduling. Slurm overcomes this issue by allowing to delay jobs with burst buffer requirements which may cause their starvation. Although this issue is not commonly visible in the current systems due to relatively low utilisation of the intermediate storage, it may become more critical in the future with further adoption of burst buffers. 

We proposed a relatively simple change to the EASY-backfilling which eliminates this inefficiency. 
In our experiments, this amended EASY-backfilling results in consistently good performance. In initial experiments, we tested a few other reasonable local heuristics, like dynamic prioritising of jobs to balance utilisation of burst buffers and processors. However, none of these approaches strongly dominated this amended EASY-backfilling. To get even better results, we had to employ a relatively more complex approach, the plan-based scheduling.
The presented plan-based scheduling not only solves the basic inefficiencies but also improves overall efficiency in terms of waiting time and bounded slowdown. Furthermore, it does not impose any requirements on resource allocation, so it may be easily generalised to other kinds of resources such as main memory (RAM), High Bandwidth Memory, GPUs, FPGAs or software licenses. However, the plan-based approach requires more changes in the cluster scheduler; and is also more computationally-intensive (although our enhancements kept the limited number of iterations of the global optimisation loop).

\section*{Data Availability Statement}
The datasets and code generated during and/or analysed during the current study are available in the Figshare repository: \url{https://doi.org/10.6084/m9.figshare.14754507}~\cite{artifact}.

\section*{Acknowledgements}
This research is supported by a Polish National Science Center grant Opus (UMO-2017/25/B/ST6/00116).

The MetaCentrum workload log \cite{10.1007/978-3-319-61756-5_5} was graciously provided by Czech National Grid Infrastructure MetaCentrum. The workload log from the KTH SP2 was graciously provided by Lars Malinowsky.
%
%
%
\bibliographystyle{splncs04}
\bibliography{publication}

\end{document}